\documentclass[useAMS,usenatbib]{mn2e}
\usepackage{url}
\usepackage{graphicx}
\usepackage{verbatim}
\usepackage[usenames]{color}
\usepackage[flushleft]{threeparttable}
\DeclareGraphicsExtensions{.pdf,.png,.jpg,.mps,.eps,.ps}
\usepackage{amsmath}
\usepackage{natbib}
\usepackage{color}
\usepackage{graphicx}
\usepackage{lscape}
\usepackage{gensymb}

\newcommand\nustar{{\it NuSTAR}}

\newcommand\ginga{{\it Ginga}}
\newcommand\sax{{\it BeppoSAX}}
\newcommand\chandra{{\it Chandra}}

\newcommand\rxte{{\it RXTE}}
\newcommand\xmm{{\it XMM-Newton}}
\newcommand\inte{{\it INTEGRAL}}

\newcommand\kev{{\rm~keV}}

\newcommand\kms{\ifmmode {\rm~km\ s}^{-1} \else ~km s$^{-1}$\fi}
\newcommand\Hunit{\ifmmode {\rm~km\ s}^{-1}\ {\rm Mpc}^{-1}
        \else ~km s$^{-1}$ Mpc$^{-1}$\fi}
\newcommand\ctssec{\ifmmode {\rm~count\ s}^{-1} \else ~count s$^{-1}$\fi}
\newcommand\ergsec{\ifmmode {\rm~erg\ s}^{-1} \else
        ~erg s$^{-1}$\fi}
\newcommand\funit{\ifmmode {\rm~erg\ s}^{-1}\;{\rm cm}^{-2} \else
        ~ergs s$^{-1}$ cm$^{-2}$\fi}
\newcommand\phflux{\ifmmode {\rm~photon\ s}^{-1}\;{\rm cm}^{-2}
        \else   ~photon s$^{-1}$ cm$^{-2}$\fi}
\newcommand\efluxA{\ifmmode {\rm~erg\ s}^{-1}\;{\rm cm}^{-2}\;{\rm
        \AA}^{-1} \else ~erg s$^{-1}$ cm$^{-2}$ \AA$^{-1}$\fi}
\newcommand\efluxHz{\ifmmode {\rm~erg\ s}^{-1}\;{\rm cm}^{-2}\;{\rm
        Hz}^{-1} \else ~erg s$^{-1}$ cm$^{-2}$ Hz$^{-1}$\fi}
\newcommand\cc{\ifmmode {\rm~cm}^{-3} \else cm$^{-3}$\fi}
\newcommand\FWHM{\ifmmode {\rm~FWHM} \else ${\rm~FWHM}$\fi}
\newcommand\Msun{\ifmmode M_{\odot} \else $M_{\odot}$\fi}
\newcommand\Lsun{\ifmmode L_{\odot} \else $L_{\odot}$\fi}

\newcommand\hbeta{\ifmmode {\rm H}\beta \else H$\beta$\fi}
\newcommand\Kalpha{\ifmmode {\rm K}\alpha \else K$\alpha$\fi}
\newcommand\nh{\ifmmode N_{\rm H} \else N$_{\rm H}$\fi}

\usepackage{graphicx}
\usepackage{url}
\usepackage{verbatim}
\usepackage{color}

\title[\nustar{} reflection spectroscopy of GX~3+1]{Study of the reflection spectrum of the bright atoll source GX~3+1 with \nustar{}}
\author[Mondal et al.]{\parbox[]{6.5in}{{Aditya S. Mondal$^{1}\thanks{E-mail: adityas.mondal@visva-bharati.ac.in}$, G. C. Dewangan$^{2}$
, B. Raychaudhuri$^{1}$ }  \\
\small
$^{1}$Department of physics, Visva-Bharati, Santiniketan, West Bengal-731235, India \\
$^{2}$Inter-University Centre for  Astronomy \& Astrophysics (IUCAA), Pune, 411007 India \\
}}

\date{\today}
\begin{document}
\maketitle
\begin{abstract}
We report on the \nustar{} observation of the atoll type neutron star (NS) low-mass X-ray binary GX~3+1 performed on 17 October 2017. The source was found in a soft X-ray spectral state with $3-70\kev{}$ luminosity of $L_\text{X}\sim3\times 10^{37}$ ergs s$^{-1}$ ($\sim 16\%$ of the Eddington luminosity), assuming a distance of 6 kpc. A positive correlation between intensity and hardness ratio suggests that the source was in the banana branch during this observation. The broadband $3-70\kev$ \nustar{} spectral data can be described by a two-component continuum model consisting of a disk blackbody ($kT_\text{disc}\sim1.8\kev{}$) and a single temperature blackbody model ($kT_\text{bb}\sim2.7\kev{}$). The spectrum shows a clear and robust indication of relativistic reflection from the inner disc which is modelled with a self-consistent relativistic reflection model. The accretion disc is viewed at an inclination of $i\simeq22\degr-26\degr$ and extended close to the NS, down to $R_\text{in}=(1.2-1.8) R_\text{ISCO}\:(\simeq6.1-9.1\,R_{g}\: \text{or}\: 14-20.5$ km) which allows an upper limit on the NS radius ($\leq13.5$ km). Based on the measured flux and the mass accretion rate, the maximum radial extension for the boundary layer is estimated to be $\sim6.3\:R_{g}$ from the NS surface. However, if the disc is not truncated by the boundary layer but by the magnetosphere, an estimated upper limit on the polar magnetic field would be of $B\leq6\times10^{8}$ G. 
\end{abstract}
\section{introduction}
 Neutron star low-mass X-ray binaries (NS LMXBs) are classified into two main groups based on their correlated spectral and timing behavior in X-rays \citep{1989A&A...225...79H}. Those are the so-called Z sources, with luminosities close to or above the Eddington luminosity ($L_\text{Edd}$) and the atoll sources, with luminosities up to $\sim0.5\:L_\text{Edd}$ \citep{2010ApJ...719..201H}. The name of Z and atoll sources is associated with the shape traced in the color-color diagram (CD). This shape can be divided into two main regions, corresponding to the X-ray state of the atoll sources. The harder one is related to the island state and the softer one is related to the banana state which can be further divided as lower banana and upper banana states. The source spectral and timing properties corresponding to the position on the CD are well determined by the basic parameters such as mass accretion rate \citep{2001ApJ...546.1107D}. X-ray bursts are frequently observed from atoll sources. The source GX~3+1 has been identified as an atoll source based on its spectral and variability properties \citep{1989A&A...225...79H}.\\

X-ray emission lines from the innermost accretion disc have been observed in different NS LMXBs \citep{2007ApJ...664L.103B, 2008ApJ...674..415C, 2008ApJ...688.1288P, 2009MNRAS.399L...1R, 2015MNRAS.451L..85D}. Disc lines are produced when the inner part of the accretion disc is illuminated by the hard X-ray emission. The hard X-ray emission could be thermal or non-thermal in nature. This process produces different spectral signatures including emission lines, absorption edges and a reflection hump that peaks between $20-30\kev{}$ \citep{2001MNRAS.327...10B, 2007MNRAS.381.1697R}. The overall interaction is known as "disc reflection''. The most prominent line produced in this process is typically an Fe K line due to its large abundance and fluorescent yield. The intrinsically narrow Fe lines appears as broad, asymmetric shape in the X-ray spectra because of the relativistic effects induced from strong gravitational field \citep{2007ARA&A..45..441M, 2000PASP..112.1145F}. This line profile is sensitive to the inner radius of the accretion disc as the relativistic effects are stronger in this area. Thus, The Fe-K emission lines are the best suited features to diagnose the accretion flows close to the NS. The accretion disc in NS systems could be truncated by the boundary layer between the disc and the NS surface or by a strong stellar magnetic field. Thus the inner disc radius sets an upper limit to the radius of the NS and hence can constrain the NS equation of state \citep{2000A&A...360L..35P, 2008ApJ...674..415C, 2011MNRAS.415.3247B}. Fe K line profile can also be used to obtain a magnetic field constraint for pulsars \citep{2009ApJ...694L..21C}. \\

GX~3+1 is one of the most luminous and persistently bright atoll sources associated with a bulge component of our Galaxy. Bright Galactic bulge X-ray sources share many common properties like a soft thermal X-ray spectrum (typically $kT\sim2-10\kev{}$), high X-ray luminosities ($10^{37}-10^{38} \text{ergs}\: s^{-1}$) and a moderate, irregular intensity variation with apparent lack of periodic behavior. These sources are also known for potential X-ray burst sources and lies always in the banana state. These bright sources have so far not shown kHz quasi-periodic oscillations (QPOs) \citep{1998ApJ...499L..41H, 2001A&A...366..138O}. However, two branch structures have been observed in the CD and hardness-intensity diagram (HID) of GX~3+1 \citep{1987MNRAS.226..383L,1998ApJ...499L..41H, 2002ApJ...568L..35M}. These two branches are identified with lower and upper banana states \citep{1993PASJ...45..801A}. The island state has so far not been observed from GX~3+1.  \\

GX~3+1 was discovered during an {\it Aerobee}-rocket flight on 1964 June 16 \citep{1965Sci...147..394B}. Being a persistent, bright X-ray source GX~3+1 has been observed with major X-ray missions, including all-sky monitor (ASM) on \ginga{} \citep{1993PASJ...45..801A}, {\it EXOSAT} \citep{1989A&A...225...48S}, \rxte{} \citep{1993A&AS...97..355B, 2000A&A...356L..45K}, \sax{} \citep{2003A&A...400..633D}, \inte{} \citep{2006A&A...459..187P}, \chandra{} \citep{2014ApJ...793..128V}, \xmm{} \citep{2015MNRAS.450.2016P} and recently with \nustar{} (present work). The source was found with a $2-10\kev{}$ luminosity of $(2-4)\times10^{37}\: \text{erg}\: \text{s}^{-1}$ \citep{2003A&A...400..633D}. The maximum persistent bolometric luminosity was found to be $\sim6\times10^{37}\: \text{erg}\: \text{s}^{-1}$. The detection of X-ray bursts confirms that it is a NS system \citep{1983ApJ...267..310M}. A unique superburst with a decay time of $1.6 \:\text{hr}$ was detected with the ASM on \rxte{} \citep{2002A&A...383L...5K}. The best distance estimate of $\sim6.1$ kpc is derived from the properties of a radius-expansion burst \citep{2000A&A...356L..45K}. The counterpart of GX~3+1 was identified with $K_{s}=15.8\pm0.1$ mag star based on the near-infrared (NIR) spectrum \citep{2014ApJ...793..128V}. Like other bright atolls GX~9+1 and GX~9+9, GX~3+1 shows strong long term X-ray flux modulations with a timescale of $\sim 6$ yr \citep{2010MNRAS.402L..16K}. \\

Spectral analysis of the source showed that its X-ray spectrum can be well described by a two-component model, consisting of a soft blackbody component, most likely associated with the accretion disc and a thermal Comptonized component/hot blackbody component which is related to the emission from the NS boundary layer \citep{2001A&A...366..138O, 2012A&A...542L..27P,2012ApJ...747...99S, 2015MNRAS.450.2016P}. Broad Fe-K emission lines around $\sim 6.5\kev{}$ have been reported in the previous \sax{}, \inte{}, \rxte{} and \xmm{} observations \citep{2001A&A...366..138O, 2012A&A...542L..27P,2012ApJ...747...99S, 2015MNRAS.450.2016P}. This feature is associated with the reflection of hard photons from the inner region of the accretion disc. \citet{2015MNRAS.450.2016P} have found that the relativistic reflection is produced at a radius of $\sim10\:R_{g}$ and the inclination angle of the system is consistent with $\sim35\degree$. However, \citet{2012A&A...542L..27P} inferred an inner disc radius $\sim25\:R_{g}$ and a disc inclination of $35\degree<i<44\degree$ during the fainter phase of the source. \citet{2019ApJ...873...99L} have also analyzed this coordinated \nustar{} observation in the $3.0-20.0$\kev{} energy band. They used double thermal model with a power-law component to fit the continuum of the soft spectrum. They detected the presence of strong reflection features. The reflection model determined the inner radius of the accretion disc which is $10.8^{+1.2}_{-3.6}\:R_{g}$ and inclination of $27\degree-31\degree$.\\

Broadband energy coverage of the \nustar{} \citep{2013ApJ...770..103H} observation of the source GX~3+1 allows us to study the source broadband spectrum and to constrain the reflection component properties such as the broad Fe emission line along with the Compton hump. These studies are important to infer the properties of the accretion flow close to the NS. In this paper, we report on a detailed study of the reflection features and the fit, with a self-consistent reflection model. In this way, this study allows us to constrain the stellar radius and/or inner accretion disk radius. We also comment on the geometry of the boundary layer between the accretion disc and the stellar surface. We organize the paper in the following way. First, we describe the observations and the details of data reduction in sec .2. In sec. 3 and sec. 4, we describe the temporal and spectral analysis, respectively. Finally, in sec.5, we discuss our findings. 

\begin{figure*}
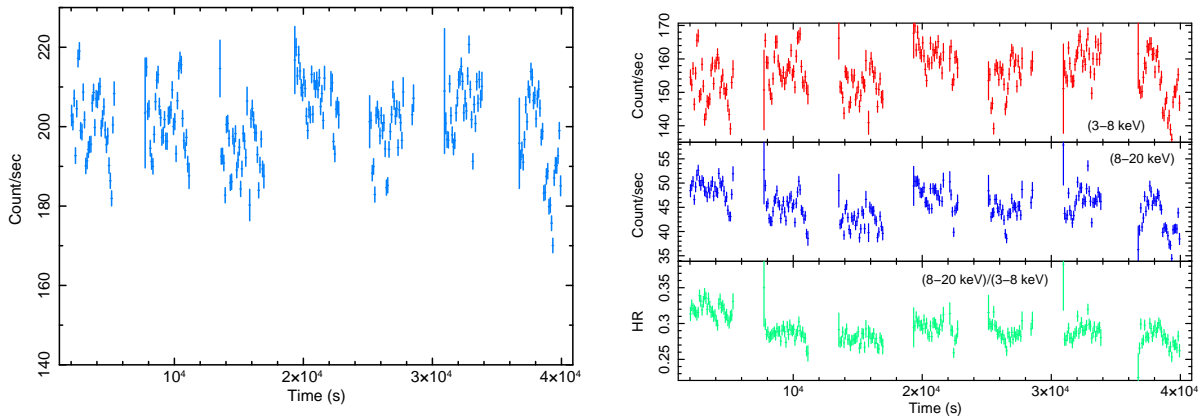

\centering
\includegraphics[scale=0.32, angle=-90]{full_lc1.ps}
\includegraphics[scale=0.32, angle=-90]{HR.ps}
\caption{Left: $3-79\kev{}$ \nustar{}/FPMA light curve of GX~3+1 with a binning of 100 sec. Right: The upper and the middle panels show the source count rate in the energy band $3-8$ keV and $8-20$ keV, respectively. Bottom panel shows the HR which is the ratio of count rate in the energy band $8-20$ keV and $3-8$ keV. } 
\label{Fig1}
\end{figure*}

\section{observation and data reduction}
\nustar{} observed the source GX~3+1 on 2017 October 17 (MJD $5804.3731$) for a total exposure time of $\sim 13.7$ ks (Obs. ID: $30363001002$). \nustar{} data of the source GX~3+1 were collected with the two co-aligned grazing incidence hard X-ray imaging Focal Plane Module telescopes (FPMA and FPMB) in the $3-79 \kev$ energy band. \\

The data were reprocessed with the standard \nustar{} data analysis software ({\tt NuSTARDAS v1.7.1}) and {\tt CALDB} ($v20181030$). We used the {\tt nupipeline} tool (version v 0.4.6) to filter the event lists. We used a circular extraction region with a radius of 100 arcsec centered around the source position to extract the source events for both the telescopes, FPMA and FPMB. We used another 100 arcsec circular region away from the source position for the purpose of background subtraction. Using the {\tt nuproducts} tool, we created lightcurve, spectra and response files for the FPMA and FPMB. We grouped the FPMA and FPMB spectral data with a minimum of 100 counts per bin and fitted the two spectra simultaneously.

\section{Temporal Analysis}
\begin{figure}
\centering
\includegraphics[width=7.0cm, angle=0]{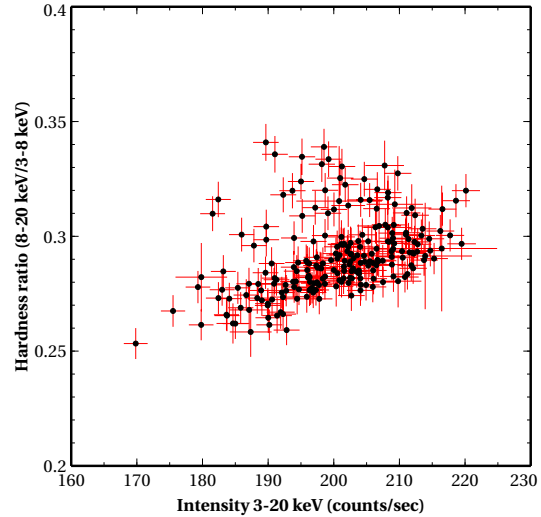}
\caption{The hardness-intensity diagram (HID) of GX~3+1. Intensity is taken as $3-20$ keV source count rate and hardness ratio has been taken as the ratio of count rate in the energy band $8-20$ keV and $3-8$ keV. The variation of hardness ratio with intensity for the whole streatch of the lightcurve is shown.} 
\label{Fig2}
\end{figure}

Left panel of Figure~\ref{Fig1} shows the $3-79\kev{}$ \nustar{}/FPMA light curve of GX~3+1 with a binning of 100 sec and spans $\sim14$ ks. The source was detected at an average intensity of $\simeq200$ counts s$^{-1}$. No X-ray bursts were observed during this observation. We also extracted the $3-8\kev{}$ and $8-20\kev{}$ light curve separately with 100~s bins and produced the corresponding hardness ratio (HR) which is displayed in the right panel of Figure~\ref{Fig1}. The HR value, which is a measure of the spectral shape, stayed quite constant at $\sim 0.28$ after 5 ks from the beginning of the observation. It suggests that the spectral shape of the source is stable during the whole span of the observation. In Figure~\ref{Fig2}, We show the hardness intensity diagram (HID), in which the HR ($3-8\kev{}$ and $8-20\kev{}$) is plotted as a function of the source intensity ($3-20\kev{}$). In this observation, we found that HR is positively correlated with intensity for this source. In the case of atoll sources, the positive correlation between the hardness and the intensity is characteristic to the banana branch \citep{1993PASJ...45..801A, 1989A&A...225...79H}. This means that the source stayed in the banana branch during this observation. Our HID and the conclusion based upon this is consistent with that of \citet{1993PASJ...45..801A} where they calculated HID with the \ginga{}/Large Area Counter (LAC) data with the almost similar definition of hardness ratio and intensity as mentioned above (see Figure 2 of \citealt{1993PASJ...45..801A}). However, it may be noted that the island state has so far not been observed from GX~3+1.

\section{spectral analysis}
We fitted the FPMA and FPMB spectra simultaneously as initial fits revealed a good agreement between these two spectra. We performed the fit over the $3-70 \kev{}$ energy band using {\tt XSPEC  v 12.9}. A constant was floated between the spectra to account for uncertainties in the flux calibration of the detectors. The constant was set 1 for FPMA and left it free for FPMB. A value of 1.02 was measured for FPMB. For each fit, we included the {\tt tbabs} model to account for interstellar absorption along the line of sight. Abundances was set to {\tt wilm} \citep{2000ApJ...542..914W} and cross-section with {\tt vern} \citep{1996ApJ...465..487V}. We fix the absorption column density to the \citet{1990ARA&A..28..215D} value of $1.16\times10^{22}$ cm$^{-2}$ as the \nustar{} data only extend down to 3 keV and found it difficult to constrain from our spectral fits. All the errors in this work are quoted at $90\%$ confidence level unless otherwise stated.  

\begin{figure*}
\includegraphics[scale=0.40, angle=-90]{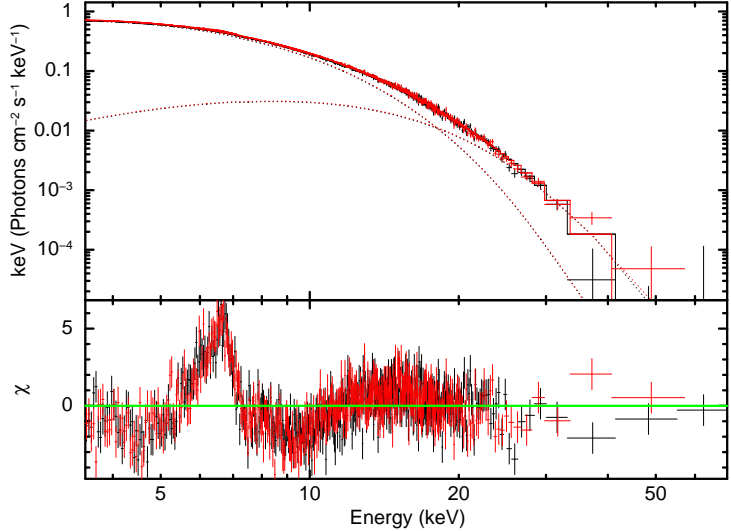}
\caption{\nustar{} (FPMA in black, FPMB in red) unfolded spectra. The spectral data were rebinned for visual clarity. Continuum is fitted with the model consisting of a multicolour disk blackbody and a single temperature blackbody. Model used: {\tt TBabs$\times$(diskbb+bbody)}. It revaled un-modelled broad emission line $\sim 5-8$ keV and a hump like feature $\sim 10-20$ keV. The prominent residuals can be indentified as a broad Fe-K emission line and the corresponding Compton hump. The red wing of the Fe-K emission line extends down to $\sim 5\kev{}$ while the blue wing drops $\sim7 \kev{}$. } 
  \label{Fig3}
   \end{figure*} 

\subsection{Continuum modeling}
For NS LMXBs in their soft states, the X-ray spectra above 7 keV are typically modelled as either a hot ($2 − 3 \kev{}$) black body or thermal Comptonization. We fitted $3-70\kev{}$ \nustar{} continuum to a model consisting of a disc blackbody component ({\tt diskbb} in {\tt XSPEC}) and a single-temperature blackbody component ({\tt bbody} in {\tt XSPEC}). This model can be simply interpreted in terms of emission from the accretion disc and boundary layer between the accretion disc and the NS surface. This combination of models gave a particularly poor fit ($\chi^{2}/dof$=$2481/850$) because of the presence of the strong disc reflection features in the spectrum which is evident in Figure~\ref{Fig3}. Emission from the boundary layer can also be modeled via low-temperature, optically thick Comptonization. To test this, we replaced the single-temperature blackbody component by the Comptonization model {\tt compTT} \citep{1994ApJ...434..570T}. But in this combination of models, the {\tt compTT} parameters are poorly constrained, particularly seed photon temperature and optical depth have taken some arbitrary large values. We note that for the earlier continuum model {\tt tbabs$\times$(diskbb+bbody)} all the continuum parameters are well constrained and thus we continued with this continuum model. We added a power-law component with the existing continuum model as this combination of spectral models, {\tt tbabs$\times$(diskbb+bbody+powerlaw)}, is also frequently used for the soft state spectra of many NS LMXBs \citep{2007ApJ...667.1073L, 2010ApJ...720..205C, 2013ApJ...779L...2M}. However, the addition of the power-law component was found to be statistically insignificant. Therefore, we proceeded with the simpler continuum model {\tt tbabs$\times$(diskbb+bbody)} as it describes the continuum fairly well and this combination of models have been widely used to fit the spectra of different NS LMXBs \citep{2010ApJ...720..205C, 2007ApJ...667.1073L}. We have shown the fitted continuum model {\tt tbabs$\times$(diskbb+bbody)} and the $\chi$ residuals in Figure~\ref{Fig3}. \\

\begin{figure*}
   \includegraphics[scale=0.40, angle=-90]{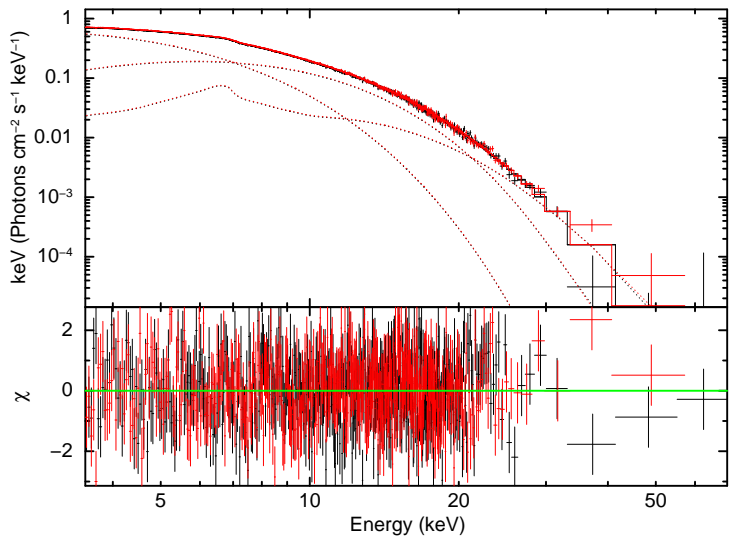}
   
 \caption{The \nustar{} (FPMA in black, FPMB in red) unfolded spectra of GX~3+1 with the best-fitting fitted model consisting of a disk blackbody, a single temperature blackbody and relativistically blurred reflection i.e.,{\tt TBabs$\times$(diskbb+bbody+relconv*reflionx)}. Lower panel shows residuals in units of $\sigma$.} 
   \label{Fig4}
   \end{figure*}

\subsection{Reflection Model}
The continuum model consisting of a disk blackbody and a single-temperature blackbody left large positive residuals around $\sim 5-8\kev{}$ and $10-20\kev{}$ (see Figure~\ref{Fig3}). The broad feature $\sim5-8\kev{}$ is consistent with Fe K$\alpha$ emission and the flux excess in the $10-20\kev$ is the corresponding Compton back-scattering hump. As these features are the clear signature of disk reflection, we proceeded by modeling our data with physical reflection models. We employed {\tt reflionx} \citep{2005MNRAS.358..211R} model which describes reflection from an ionized disc. \\

Our broadband continuum fits prefer a blackbody model over a Comptonized model to describe the spectrum at higher energies. Moreover, it is clear in Figure~\ref{Fig3} that most of the flux capable of ionizing Fe comes from the blackbody component. We therefore included a modified version of the {\tt reflionx} \footnote{https://www-xray.ast.cam.ac.uk/\~{}mlparker/reflionx\_models \\/reflionx\_bb.mod} model that assumes the disc is illuminated by a blackbody, rather than a power law (see e.g. \citealt{2010ApJ...720..205C, 2016ApJ...819L..29K, 2016MNRAS.456.4256D}). The parameters of the {\tt reflionx} model are as follows: the disc ionization parameter ($\xi$), the iron abundance ($A_{Fe}$), the temperature of the ionizing black body flux $kT_{refl}$ and a normalization $N_{refl}$. We convolved {\tt reflionx} with {\tt relconv} \citep{2010MNRAS.409.1534D} in order to account for relativistic Doppler shifts and gravitational redshifts. The emissivity of the disk in the model {\tt relconv} is described as a broken powerlaw in radius (e.g., $\epsilon\propto r^{-q}$), giving three parameters: inner emissivity index ($q_{in}$), outer emissivity index ($q_{out}$) and break radius ($R_{break}$). Here we used a constant emissivity index (fixed slope) by fixing $q_{out}=q_{in}$ (obviating the meaning of $R_{break}$) as the slope is not constrained by the data. The fit parameters of the {\tt relconv} model are as follows: the emissivity index ($q$), the inner and outer disk radius $R_{in}$ and $R_{out}$, the disk inclination ($i$) and the dimensionless spin parameter ($a$). \\

We introduced a few reasonable conditions when making fits with {\tt reflionx} and {\tt relconv}. We set the emissivity to $q=3$, in agreement with a Newtonian geometry far from the NS \citep{2010ApJ...720..205C}. Following \citet{2000ApJ...531..447B}, the dimensionless spin parameter $a$ can be approximated as $a\simeq0.47/P_{ms}$ where $P_{ms}$ is the spin period in ms. But the spin period of the source GX~3+1 is not known. The fastest known NSs spin at $\sim 1.5$ ms which corresponds to $a\simeq0.3$ \citep{2008ApJS..179..360G}. The innermost stable circular orbit (ISCO) is then located at $R_{ISCO}=5.05\,R_{g}$, where $R_{g}=GM/c^2$ is the gravitational radius \citep{2016MNRAS.461.4049D}. For $a=0$ the position of the ISCO is at $R_{ISCO}=6\,R_{g}$. Thus there is a small shift in the position of the ISCO compared to the Schwarzschild metric ($a=0$). We performed the fit with $a=0$ as well as $a=0.3$. We note that both the fit yielded similar results as expected. We also fixed the outer radius $R_{out}=1000\;R_{g}$. Further, we fixed the $A_{Fe}$ to unity (compatible with the solar value) as the fit was almost insensitive to this parameter.\\   

The addition of the relativistic reflection model improved the spectral fits significantly ($\chi^{2}/dof$=$899/845$). The best-fit parameters for the continuum and the reflection spectrum are shown in Table~\ref{parameters}. Our fits suggest that the inner disc is located close to the NS at $R_{in}=(1.2-1.8) R_{ISCO} (\simeq6.1-9.1\,R_{g}\: \text{or}\: 14-20.5$ km). The inclination angle is found to be $i=24\pm2$ degree in agreement with the fact that neither dips nor eclipses have been observed in the light curve of GX~3+1. The reflection component has an intermediate disc ionization of $\xi\simeq151-236$ erg s$^{-1}$ cm which is the typical range observed in both black holes and NS LMXBs ($\text{log}\xi\sim 2-3)$. The fitted spectrum with relativistically blurred reflection model and the residuals are shown in Figure~\ref{Fig4}. \\

\begin{figure*}
\includegraphics[width=7.0cm, angle=0]{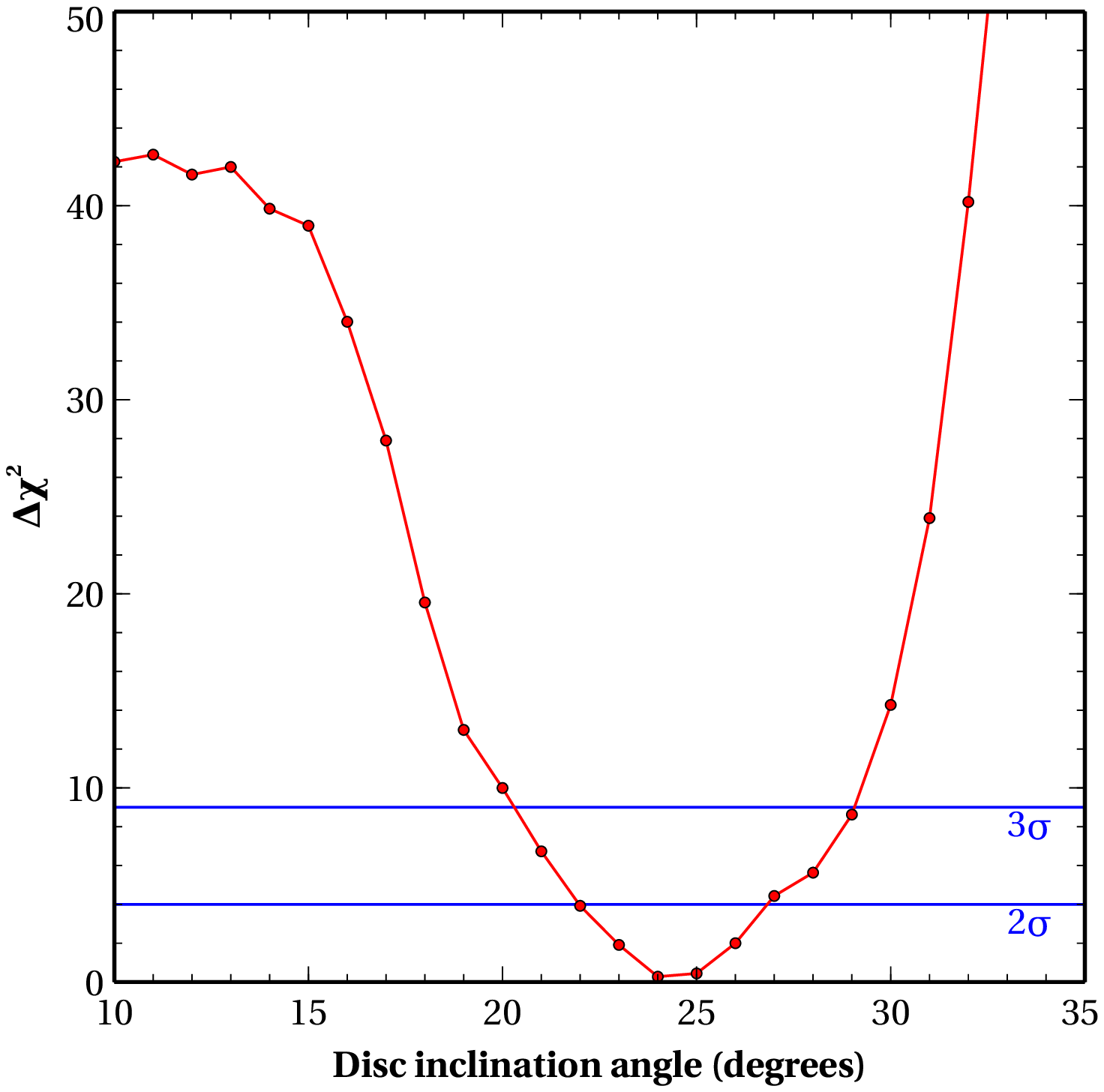}\hspace{1cm}
\includegraphics[width=7.0cm, angle=0]{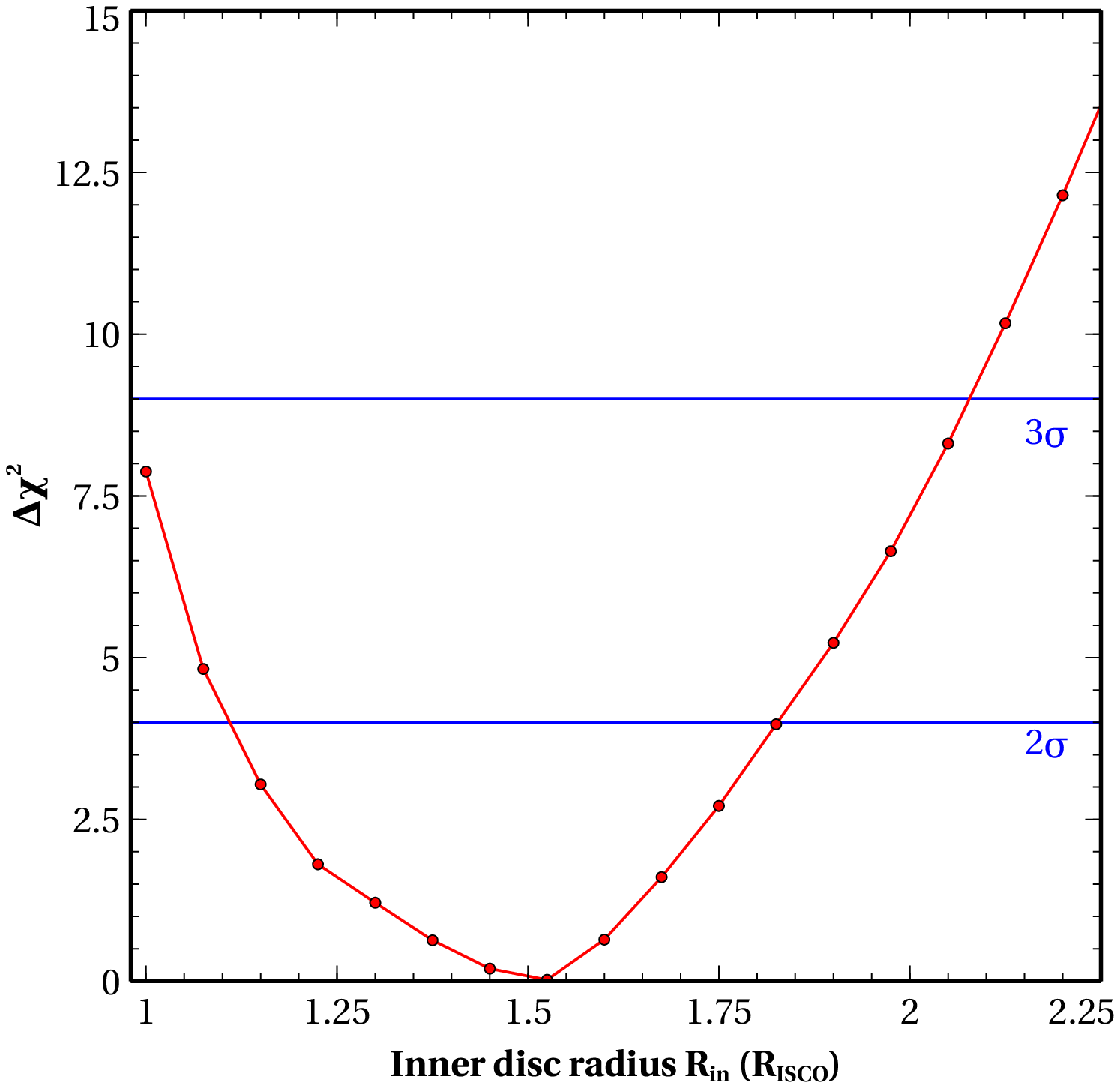}
\caption{left panel shows the variation of $\Delta\chi^{2}(=\chi^{2}-\chi_{min}^{2})$ as a function of disk inclination angle obtained from the relativistic reflection model ({\tt reflionx$\_$bb.mod}). We varied the disc inclination angle between 10 degree and 35 degree. Right panel shows the variation of $\Delta\chi^{2}(=\chi^{2}-\chi_{min}^{2})$ as a function of inner disc radius (in the unit of $R_{ISCO}$) obtained from the relativistic reflection model. We varied the inner disc radius as a free parameter upto $2.5\,R_{ISCO}$. Horizontal lines in both the panels indicate $2\sigma$ and $3\sigma$ significance level.} 
\label{Fig5}
\end{figure*}

In order to constrain the inner disc radius and the disc inclination angle from our best-fit model, we computed $\Delta\chi^2$ for each of the parameters using {\tt steppar} in {\tt xspec}. Figure~\ref{Fig5} shows plots of $\Delta\chi^2$ versus the disc inclination angle and the inner disc radius for the best-fit model in the left and right panel, respectively. This figure manifests the sensitivity of the spectrum to the inner extent of the disc as well as to the disc inclination angle. It strongly prefers a disc that is close to the ISCO and is statistically consistent with the disc extending to the ISCO itself. \\

Considering the fact that the reflection parameters are mostly constrained by the iron line, we tried to fit the data with the {\tt relline} model, a relativistic line profile excluding the broadband features such as the Compton hump. We obtained the line energy $E$ equal to $7.00^{+0.02}_{-0.07}$\kev{}. The value for the inner disc radius, $R_{in}=1.36^{+0.11}_{-0.09}\; R_{ISCO}$ is consistent with our above estimation. We found a small inclination of $\sim 15\degree$. However, the {\tt relline} model, with $\chi^{2}/dof$=$1042/846$, is not as good a fit as the broadband reflection model described above ($\chi^{2}/dof$=$899/845$). It suggests that the broadband reflection spectrum does make a significant contribution. \\

Additionally, we also tried to fit the spectrum with another flavor of the {\tt reflionx} \citep{2005MNRAS.358..211R} model that assumes reflection of a power-law with a high energy exponential cutoff. To take relativistic blurring into account, we convolved {\tt reflionx} with {\tt relconv} \citep{2010MNRAS.409.1534D}. We continued our analysis with the reasonable values of some parameters mentioned above but the $A_{Fe}$ left free. This model significantly worsens the fit, resulting in a $\chi^{2}/dof$=$1040/844$, yet it does not cause a large change to the main parameter of interest, $R_{in}\;(2.16^{+0.52}_{-0.61}\; R_{ISCO})$ and inclination $i\;(\sim27\degree)$. This fit tended towards the larger value of $A_{Fe}\ge9.16$. This fit led to the smaller blackbody temperature $(kT_{bb})$ compared to the disc temperature $(kT_{disc})$ which is quite unphysical. The problem may lie in the difference of the shape of the reflection spectrum which assumes an input {\tt power-law} to a reflection spectrum that assumes a {\tt blackbody} input spectrum. Moreover, When we performed the fit again forcing $A_{Fe}$ equal to the solar value, all the important parameters become unconstrained and assume unphysical values. \\

This overabundance of iron could be indicative of a higher density disc \citep{2018ApJ...855....3L}. It may be noted that the model {\tt relxillD} provides the option of variable density in the disk. Therefore, we applied a cut-off power-law reflection model with variable disc density to account the very high iron abundances implied in the reflection fitting with {\tt reflionx}. This model, {\tt tbabs$\times$(diskbb+bbody+relxillD)} provided $\chi^{2}/dof$=$950/842$. Although, we were unable to constraint some important parameters like reflection fraction ($refl_{frac}$) and normalization but it did not change the results much for important parameters, i.e., $R_{in}\;(\le1.5\; R_{ISCO})$ and inclination $i\;(\sim25\degree)$. The density of the disc was found to be high, $\text{log}\:n\:(\text{cm}^{-3})=18.1\pm0.1$. Although, this fit led to an increment of the reduced $\chi^{2}$ still we are unable to draw any conclusion based on this model as some important parameters become unconstrained. Moreover, the current version of this model has a fixed cutoff energy of $300$\kev{} which is much higher than the value required to fit these data \citep{2019ApJ...873...99L}. Therefore, a cutoff power-law reflection model with variable cutoff energy may be useful to serve this purpose. 

\begin{table}
   \centering
   \caption{ Best-fit spectral parameters of the \nustar{} observation of the source GX~3+1 using model: {\tt TBabs$\times$(diskbb+bbody+relconv*reflionx)}.} 
\begin{tabular}{|p{1.6cm}|p{4.0cm}|p{1.7cm}|}
    \hline
    Component     & Parameter (unit) & Value \\
    \hline
    {\scshape tbabs}    & $N_{H}$($\times 10^{22} cm^{-2}$) &$1.16(f)$     \\
    {\scshape diskbb} & $kT_{disc}$($\kev$) & $1.53_{-0.08}^{+0.13}$   \\
    & $N_{diskbb}$[(km/10 kpc)$^{2}$cos$i$]   & $58_{-12}^{+11}$    \\

    {\scshape bbody} & $kT_{bb}(\kev)$ & $2.03\pm0.08$    \\
    & $N_{bb}$~$^a$ ($\times 10^{-2}$) & $3.44\pm0.45$     \\ 
   
   {\scshape relconv} & $i$ (degrees) & $24\pm2$    \\
   & $R_{in}$($\times R_{ISCO}$) & $1.51_{-0.33}^{+0.24}$ \\
   {\scshape reflionx} & $\xi$(erg cm s$^{-1}$) & $191_{-40}^{+45}$  \\
   & $kT_{refl}$ (keV) & $2.60_{-0.08}^{+0.10} $  \\
   & $N_{refl}$~$^b$ &  $2.48_{-0.49}^{+1.22}$ \\
   & $F^{*}_{total}$ ($\times 10^{-8}$ ergs/s/cm$^2$) & $0.66\pm 0.01$ \\
   & $F_{diskbb}$ ($\times 10^{-8}$ ergs/s/cm$^2$) & $0.31 \pm 0.01$  \\
   & $F_{bbody}$ ($\times 10^{-8}$ ergs/s/cm$^2$)& $0.28 \pm 0.01$ \\
   & $F_{reflionx}$ ($\times 10^{-8}$ ergs/s/cm$^2$) & $0.07 \pm 0.01$  \\
   & $L_{3-79 keV}$ ($\times 10^{37}$ ergs/s) & $2.84 \pm 0.01$  \\	
  
\hline 
    & $\chi^{2}/dof$ &$899/845$   \\
    \hline
  \end{tabular}\label{parameters} \\
 
The outer radius of the {\tt relconv} spectral component was fixed to $1000\;R_{G}$ and the spin parameter was set to $a=0.3$ and $q=3$. Iron abundance ($A_{Fe}$) was set to 1. ~$^{a,b}$ denotes the normalization component of the {\tt bbody} and {\tt relconv} model, respectively. Assumed a distance of 6 kpc and a mass of $1.5\;M_{\odot}$ for calculating the luminosity. $^{*}$All the fluxes are calculated in the energy band $3.0-79.0\kev$. 
\end{table}

\begin{figure*}
\includegraphics[width=6.0cm, angle=0]{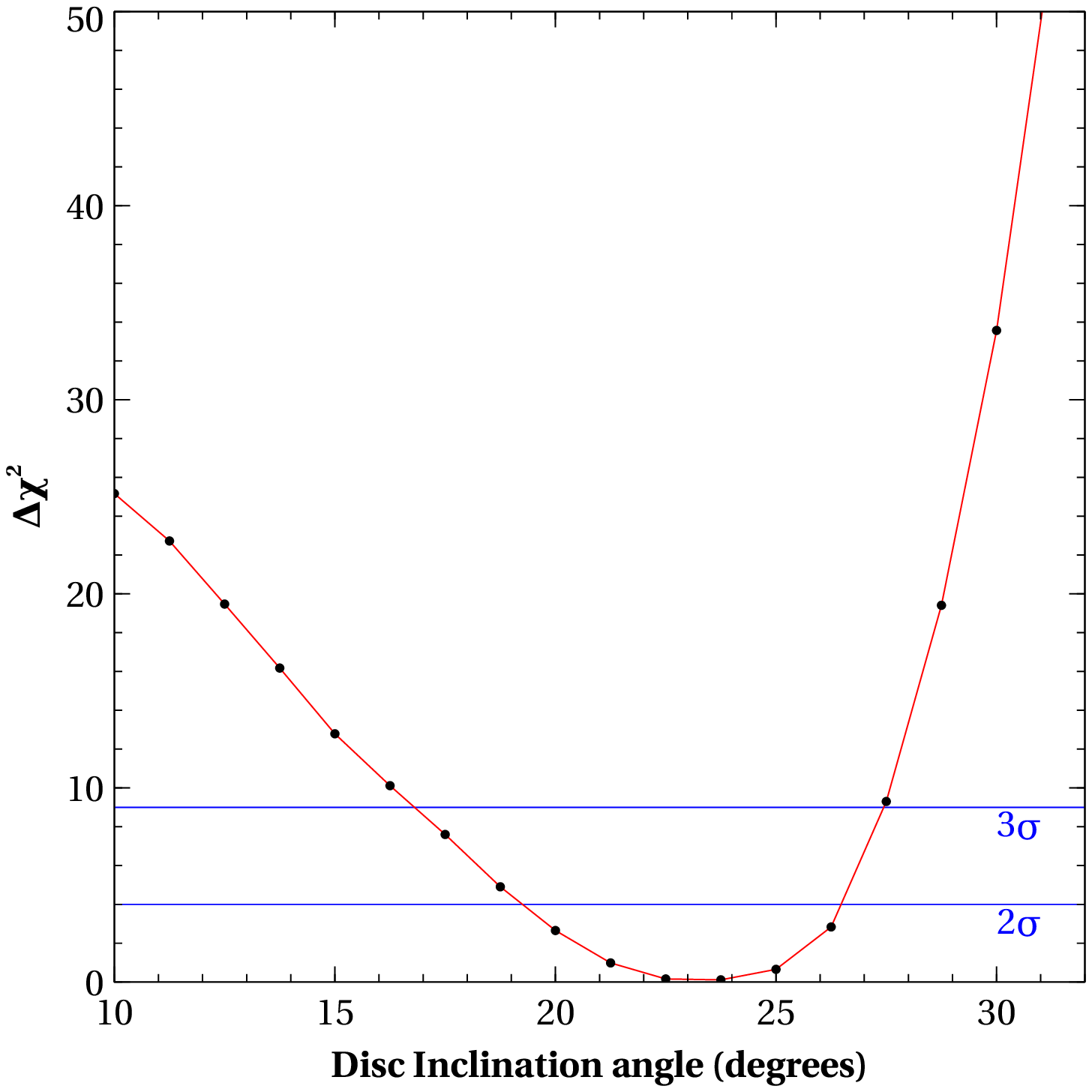}\hspace{1cm}
\includegraphics[width=6.0cm, angle=0]{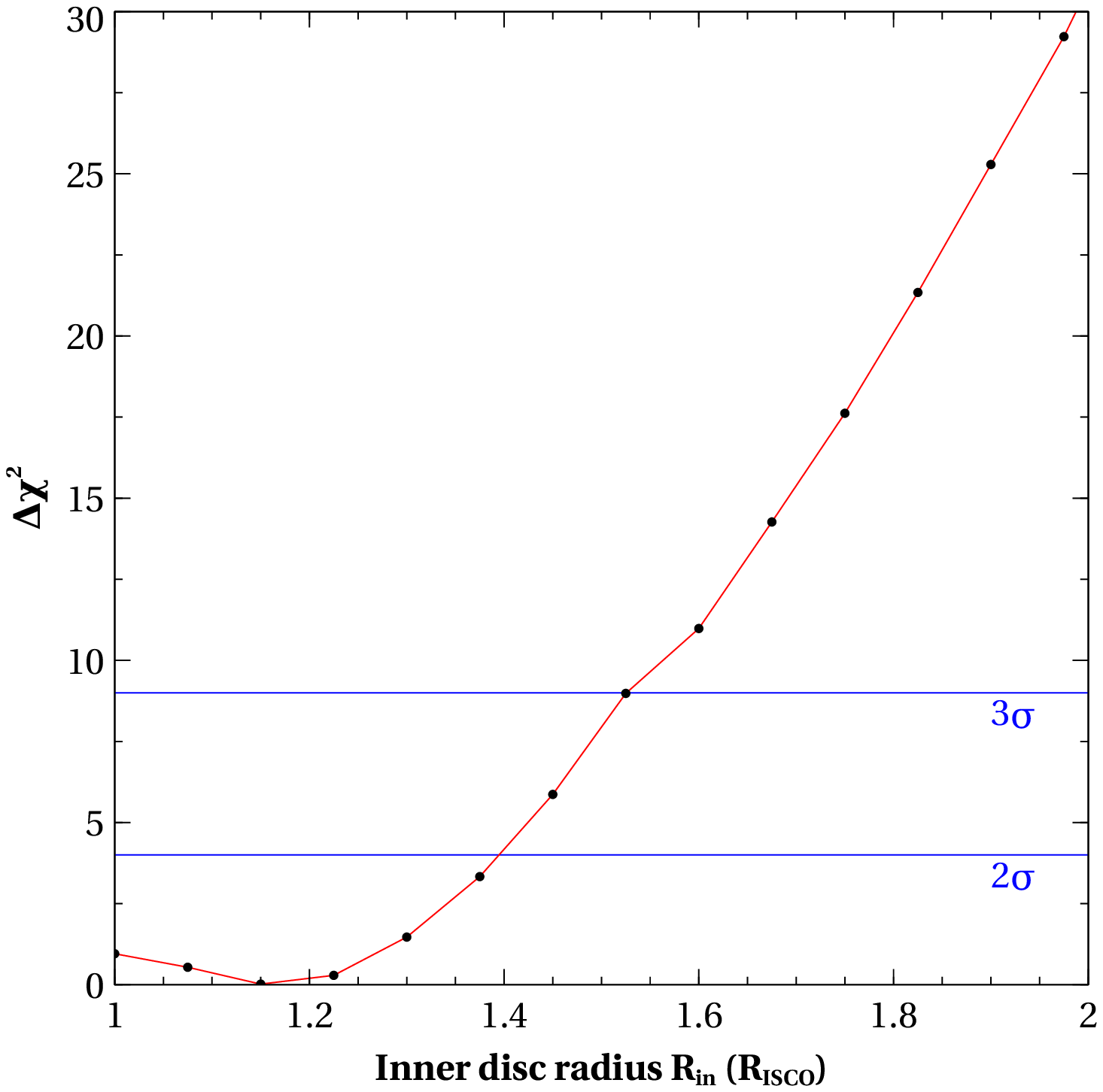}\hspace{1cm}
\includegraphics[width=6.0cm, angle=0]{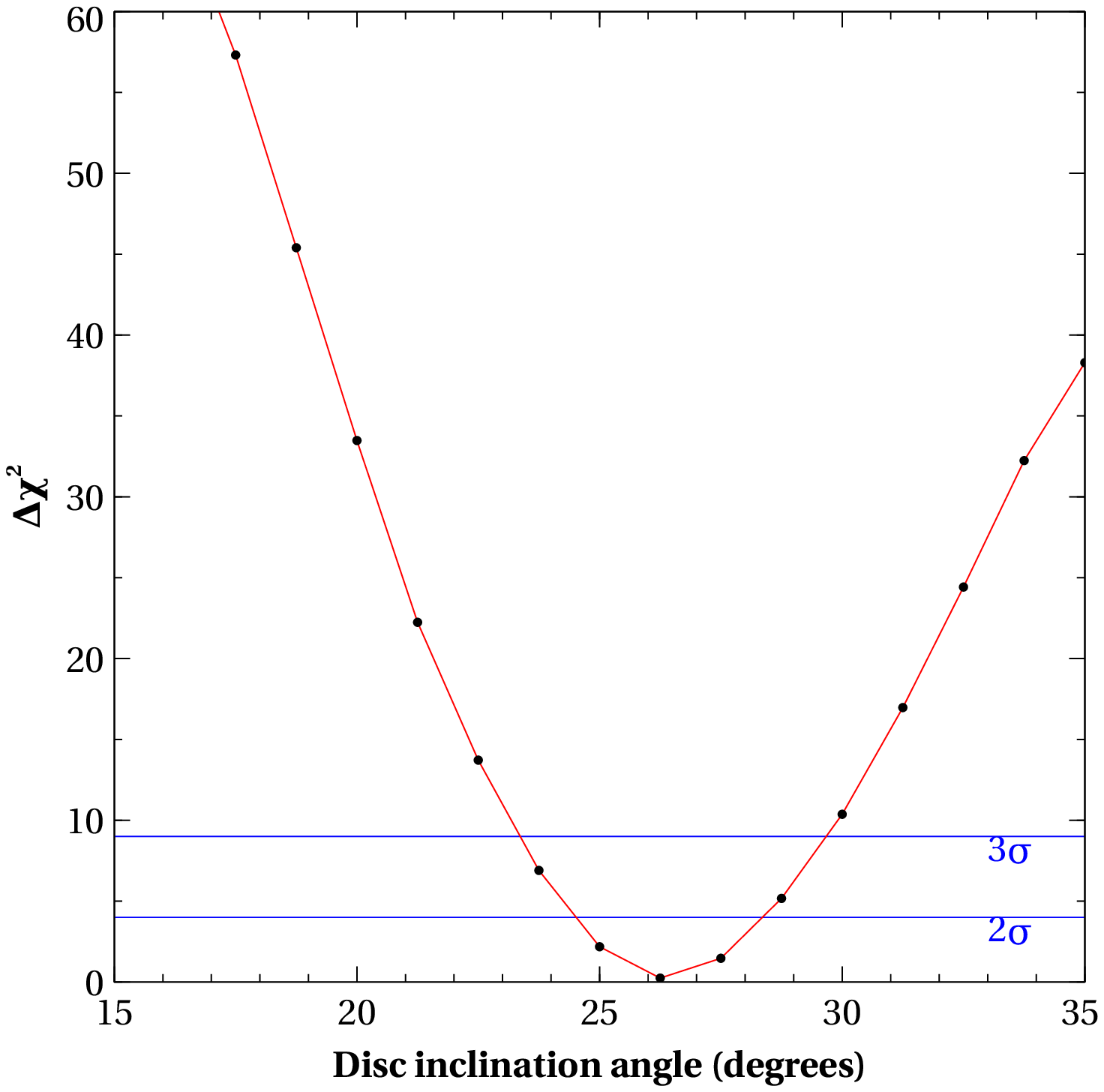}\hspace{1cm}
\includegraphics[width=6.0cm, angle=0]{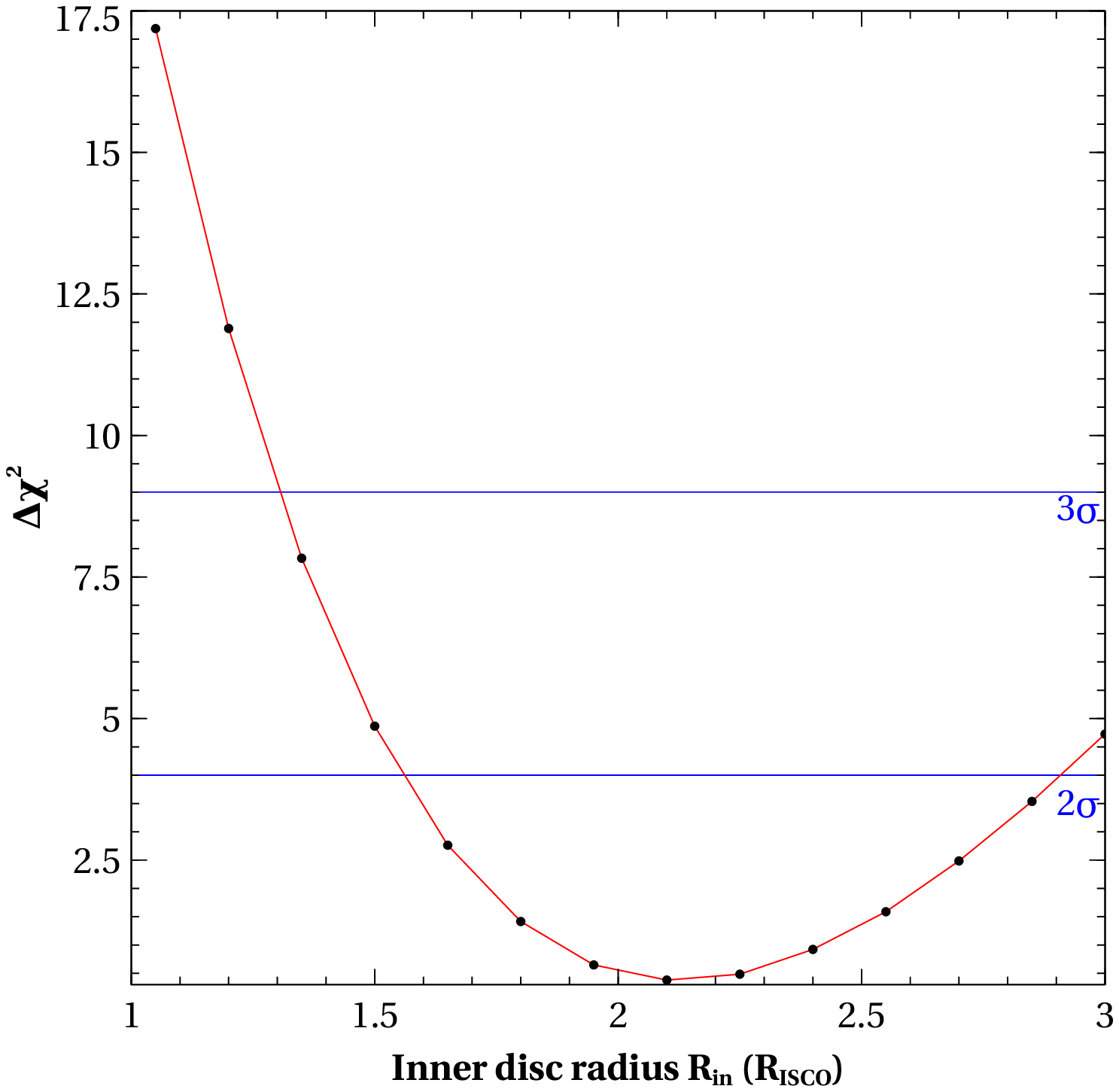}
\caption{Contour plots, same as Figure~\ref{Fig5}, but for other relativistic reflection models considered during analysis. First two plots show the variation of $\Delta\chi^{2}(=\chi^{2}-\chi_{min}^{2})$ as a function of disk inclination angle and inner disc radius (in the unit of $R_{ISCO}$) respectively, obtained from the relativistic reflection model {\tt relxillD}. Last two plots show the same for the relativistic reflection model {\tt reflionx} which assumes {\tt power-law} as an input spectrum.}
\label{Fig6}
\end{figure*}

\section{Discussion}
We report on the \nustar{} observation of the bright atoll type NS LMXB GX~3+1. The source was in a soft spectral state with the $3-70\kev$ luminosity of $L_{X}\sim2.84\times 10^{37}$ ergs s$^{-1}$, assuming a distance of 6 kpc. This corresponds to $\sim 16\%$ of the Eddington luminosity, confirming predictions from the theoretical modelling of the X-ray spectra of bright sources like GX~3+1 \citep{1998AIPC..431..125P}. From the hardness-intensity diagram (HID), it is confirmed that the source was in the so-called banana branch during the psesent observation. The broad-band $3-70\kev$ \nustar{} spectral data can be described by a continuum model consisting of a disk blackbody {\tt diskbb} ($kT_{disc}\sim1.8\kev{}$) and a single temperature blackbody {\tt bbody} model ($kT_{bb}\sim2.7\kev{}$). Thermal emission from the accretion disc is prominently detected in the X-ray spectrum. The hot blackbody emission provides most of the hard X-ray flux that illuminates the accretion disc and produces the reflection spectrum. The spectral data required a significant reflection component, characterized by the broad Fe-K emission line $\sim 6-7\kev{}$ and a Compton hump around $10-20\kev{}$. Studying reflection spectra provides valuable insight into the accretion geometry, such as the inner radius of the accretion disc, inclination of the accretion disk and height of the illuminating X-ray source. \\

\citet{2019ApJ...873...99L} have also analyzed this coordinated \nustar{} observation. They modelled the \nustar{} data in the $3.0-20.0$\kev{} energy band. To fit the continuum, they used double thermal model ({\tt diskbb} and {\tt bbody}) with a {\tt power-law} component. They also detected the presence of strong reflection features which are properly described by the self-consistent thermal reflection model {\tt RELXILLNS}. This reflection model assumes a blackbody is irradiating the accretion disc. In their fitting, replacing of the {\tt diskbb} and {\tt power-law} components with {\tt nthcomp} led to the improvement of the overall fit but causing a high optical depth ($\tau\sim7$). They also tested another reflection model {\tt RELXILLCp}, that allows for reflection from an {\tt nthcomp} Comptonization continuum. But this model provided a significantly worse fit compared to the previous one and they did not report on it. They measured the emissivity profile of the accretion disc which was found to be consistent with a single unbroken power-law with index $q=3.2^{+0.1}_{-0.6}$. This is also consistent with our work where a single emission line ($q=3$) is fitted over the entire disc. Their best-fit reflection model determined the inner radius of the accretion disc which is $1.8^{+0.2}_{-0.6}\:R_{ISCO}\;(10.8^{+1.2}_{-3.6}\:R_{g})$ and inclination of $27\degree-31\degree$.\\

A characteristic reflection spectrum is produced where the accretion disc is illuminated by an external X-ray source which could be the emission generated in a hotspot on the surface of the NS or in the boundary layer or emission from a corona associated with the disc. \citet{2018MNRAS.475..748W} studied on the illumination of the accreting NS X-ray binaries to understand the nature of the primary X-ray source that illuminates the disc through the study of the emissivity profile. A hotspot on the surface of the NS would likely be created when the accreting material is channeled down along the magnetic field lines on to the pole of the dipole field. Such a strong magnetic field is expected to truncate the accretion disc at a larger inner disc radius. Since no such significant disc truncation is observed, it follows that the accreting material is not directed by the magnetic field to form the hotspots on the NS surface. In one hand, there is no evidence for a significant contribution to the non-thermal X-ray emission arising from a corona associated with the accretion disc. On the other hand, there are potential evidences on the fact that the disc is predominantly illuminated by X-rays emitted from close to the NS surface itself which could be identified as the emission from the boundary layer between the NS surface and the inner part of the accretion disc.\\

\subsection{Inner radius and the inclination of the accretion disc}
We fitted the reflection features with a relativistically blurred reflection model {\tt reflionx} which uses blackbody input spectrum to investigate the accretion geometry of GX~3+1. It is generally believed that in the soft X-ray spectral state when the luminosity of the source is $\geq10\%$ of the Eddington limit, the accretion disc extends to/near the ISCO (e.g. \citealt{1997ApJ...489..865E}). We found that the inner edge of the accretion disc extended inwards to $R_{in}=(1.2-1.8)\:R_{ISCO}$, which is consistent with \citet{2019ApJ...873...99L}. Given that $R_{ISCO}\simeq5.05\:GM/c^2$ for a NS spinning at $a=0.3$, this would correspond to $R_{in}=(6.1-9.1)\,R_{g}$ or $(14-20.5)$ km for a $1.5M_{\odot}$ NS. The inferred inner disc radius is also consistent with high luminosity implied during this observation. Similar inner disc radius is obtained by \citet{2017ApJ...847..135L} and \citet{2016MNRAS.461.4049D} when they analyzed the \nustar{} spectra of NS LMXB Aquila~X-1 and 1RAX~$J180408.9-342058$, respectively in their soft spectral states. Moreover, our estimated inner disc radius for GX~3+1 is within the range obtained for several other NS LMXBs ($R_{in}\simeq5-20 GM/c^2$; see \citealt{2010ApJ...720..205C, 2015MNRAS.451L..85D, 2015MNRAS.449.2794D, 2011ApJ...731L...7M, 2013MNRAS.429.3411P}). \\

We compared the inner disk radius from the Fe line fitting with that implied from the {\tt diskbb} fits. The normalization component of the {\tt diskbb} is defined as $N_{diskbb}=(R_{in, diskbb}/D_{10})^{2}$cos$i$ (where $R_{in, diskbb}$ in km and $D_{10}$ is the distance in units of 10 kpc), which can be used as an important probe to constrain the inner radial extent of the accretion disk. Different correction factors such as inner boundary assumptions \citep{1999MNRAS.309..496G}, spectral hardening \citep{2000MNRAS.313..193M}, absolute flux calibration of the instrument or the spectral model need to be taken into account to calculate the true inner-disk radii from {\tt diskbb} fits. From our best fit {\tt diskbb} normalization, we obtained an inner disk radius of $R_{in,diskbb}\simeq 4.3-5.3$ km for an inclination of $i=22-26$ degree and distance of 6 kpc. However, if we corrected it with the hardening factor $\simeq1.7$ (e.g. \citealt{2001ApJ...547L.119K, 2013ApJ...769...16R}), then it yielded $R_{in,diskbb}\simeq (12.5-15.4)$ km. This is consistent with the location of the inner disc inferred from our reflection fits ($R_{in}=14-20.5$ km). \\

We found that the inner disc has a relatively low viewing angle ($i\sim24$ degree), comparable with \citet{2019ApJ...873...99L}. It is consistent with the fact that neither dips nor eclipses have been observed in the light curve of GX~3+1. This lower value of the disc inclination is also consistent with \citet{2015MNRAS.450.2016P} and \citet{2012A&A...542L..27P}.\\

We have further tested two other reflection models. One is the {\tt relxillD} which provides the option of variable density in the disc and the other is a different flavor of {\tt reflionx} which uses a cutoff {\tt power-law} input spectrum. It is interesting and important to note that we have found consistent results for the inner disk radius ($R_{in}$) and inclination ($i$) which are the parameters of fundamental interest with all three reflection models. It demonstrates that these results are not strongly affected by the assumed density or input spectral shape. This is particularly important for NSs, since the reflection models used in the past frequently assumed {\tt power-law} spectra and standard low densities. It is also interesting that the {\tt relline} model does not return the same parameters. This presumably implies that some part of the complete reflection spectrum is required to measure $R_{in}$ or $i$ reliably, but it is not sensitive to the continuum.

\subsection{Mass accretion rate}
From the persistent flux ($F_{p}$) and the distance ($d$) of the source, we can also estimate the accretion rate ($\dot{m}$) per unit area at the NS surface \citep{2008ApJS..179..360G}. Here we used equation.2 of \citet{2008ApJS..179..360G}
\begin{equation}
\begin{split}
\dot{m}=&\:6.7\times 10^{3}\left(\frac{F_{p}c_\text{bol}}{10^{-9} \text{erg}\: \text{cm}^{-2}\: \text{s}^{-1}}\right) \left(\frac{d}{10 \:\text{kpc}}\right)^{2} \left(\frac{M_\text{NS}}{1.4 M_{\odot}}\right)^{-1}\\
 &\times\left(\frac{1+z}{1.31}\right) \left(\frac{R_\text{NS}}{10\:\text{km}}\right)^{-1} \text{g}\: \text{cm}^{-2}\: \text{s}^{-1},
 \end{split} 
\end{equation}
where $c_{bol}$ is the bolometric correction which is $\sim 1.38$ for the nonpulsing sources \citep{2008ApJS..179..360G}. $M_{NS}$ and $R_{NS}$ are the NS mass and radius, respectively. $z$ is the surface redshift and $1+z=1.31$ for a NS with mass 1.4 $M_{\odot}$ and radius 10 km. We determine the mass accretion rate using $F_{p}=0.66\times 10^{-8}$ ergs s$^{-1}$ cm$^2$ to be $\sim5.1\times10^{-9}\;M_{\odot}\;\text{y}^{-1}$ during this observation. This inferred value of $\dot{m}$ is consistent with \citet{2003A&A...400..633D} and \citet{2000A&A...356L..45K}. Moreover, it is consistent when the source is in the banana branch.

\subsection{Geometry of the boundary layer}
Our upper limits on $R_{in}$ suggests that the disc is truncated substantially above the NS surface itself. It allows us to consider that the disc may be truncated by a boundary layer extending from the stellar surface \citep{2017ApJ...847..135L}. Equation (2), given by \citet{2001ApJ...547..355P}, provides a way to estimate the maximum radial extent ($R_\text{max}$) of the boundary layer region from the mass accretion rate.
\begin{equation}
\text{log}(R_\text{max}-R_\text{NS})\simeq5.02+0.245\left|\text{log}\left(\frac{\dot{m}}{10^{-9.85}\:M_{\odot}\:\text{yr}^{-1}}\right)\right|^{2}
\end{equation}
We determine the mass accretion rate to be $\sim5.1\times10^{-9}\;M_{\odot}\;\text{y}^{-1}$ during this observation. This gives a maximum radial extent of $\sim 6.3\:R_{g}$ for the boundary layer (assuming $M_{NS}=1.5\:M_{\odot}$ and $R_{NS}=10$ km). Similar value ($\sim6.6\:R_{g}$) for the radial extent of the boundary layer is also estimated by the \citet{2019ApJ...873...99L}. This is consistent with the location of the inner disc radius measured from spectral modelling. Similar radial extent of the boundary layer also found by different authors (see \citealt{2016ApJ...819L..29K, 2017ApJ...847..135L}).  \\

We now further examined the conception that the boundary layer is the source of ionizing flux. We determined the maximum height of the boundary layer for a disc extending close to the NS surface. Following Equation (6) of \citet{2010ApJ...720..205C}, the height of the ionizing source above the disc is defined as 
\begin{equation}
Z^{2}=\frac{L_\text{BL}}{n_{e}\xi}-R_{in}^{2},
\end{equation} 
where $L_\text{BL}$ is the boundary layer luminosity, $n_{e}$ is the electron density, $\xi$ is the ionization parameter and $R_{in}$ is the inner accretion disc radius. From spectral fitting and reflection modelling we determine $L_\text{BL}$, $\xi$ and $R_{in}$. We estimated $n_{e}$ of the accretion using the relation $\displaystyle\xi=L_{X}/(n_{e}r^{2})$. Since we find that $L_{X}\sim2.84\times 10^{37}$ ergs s$^{-1}$, we obtain $n_{e}\sim 0.51\times10^{23}\:\text{cm}^{-3}$ following $r=17$ km and $\xi=191$ erg cm s$^{-1}$ (best-fit values). Thus, for $\xi=(151-236)$ erg cm s$^{-1}$, $L_\text{BL}\simeq1.28\times10^{37}\:\text{erg}\:\text{s}^{-1}$ and $R_{in}=(14-20.5)\:\text{km}$, we find $Z=(9.5-15)$ km which is equivalent to $(4.2-6.8\:R_{g}$). The small height of the ionizing source inferred from our reflection fit could refer to the boundary layer between the accretion disc and the NS surface as the primary source of the illuminating hard X-rays (see \citealt{2013MNRAS.432.1144S,2015MNRAS.451L..85D}).  

\subsection{NS radius constraints}
If the disc extends closer to the surface of the NS, then the reflection modelling can be used to place constraints on the NS radius. Reflection modelling permit us to put a lower limit on the gravitational redshift from the NS surface. Gravitational redshift is given by $\displaystyle 1+z=1/\sqrt{(1-2GM/R_\text{in}c^{2}}$. For $R_{in}=1.5\:R_\text{ISCO}\:(\simeq7.6\:GM/c^{2})$, implied by our reflection fit would constrain the NS radius to $R_\text{NS}\leq17$ km, hence the gravitational redshift to $z\geq0.16$ for an assumed mass of $M=1.5\:M_{\odot}$. Our measurement for $R_{in}$ does extend down to $1.2\:R_\text{ISCO}$. If this were the case, then it would constrain the NS radius to $R_\text{NS}\leq13.6$ km for the gravitational redshift $z\geq0.22$. This constraint on the radius of the NS is consistent with the result obtained from the analysis of the type-I X-ray bursts \citep{2003A&A...400..633D}.

\subsection{Magnetic field strength}
The inner part of the accretion disc may have also been truncated by the associated magnetic field of the NS. We can thus use our measured inner disc radius from the reflection fit to estimate an upper limit of the magnetic field strength of the NS. We used the following equation of \citet{2009ApJ...694L..21C} which was a modified version of the formulation of \citet{2009MNRAS.400..492I} to calculate the magnetic dipole moment. \\
\begin{equation}
\begin{split}
\mu=&3.5\times 10^{23}k_{A}^{-7/4} x^{7/4} \left(\frac{M}{1.4 M_{\odot}}\right)^{2}\\
 &\times\left(\frac{f_{ang}}{\eta}\frac{F_{bol}}{10^{-9} \text{erg}\: \text{cm}^{-2}\: \text{s}^{-1}}\right)^{1/2}
 \frac{D}{3.5\: \text{kpc}} \text{G}\; \text{cm}^{3}
\end{split} 
\end{equation}
where $\eta$ is the accretion efficiency in the Schwarzschild metric, $f_{ang}$ is the anisotropy correction factor. The coefficient $k_{A}$ depends on the conversion from spherical to disk accretion (numerical simulation suggest $k_{A}=0.5$ whereas theoretical model predict $k_{A}<1.1$). \citet{2009ApJ...694L..21C} modified $R_{in}$ as $R_{in}=x\:GM/c^{2}$. We estimated flux in the $0.01-100\kev$ range (extrapolating \nustar{} spectral fit) is of $F_{bol}=7.97\times10^{-9}$ erg cm$^{-2}$ s$^{-1}$. We assumed $D=6$ kpc, $M=1.5M_{\odot}$ and $R=10$ km. Using $R_{in}\leq9\, R_{g}$ from the \nustar{} spectral fit, along with the assumptions $k_{A}=1$, $f_{ang}=1$ and $\eta=0.1$, leads to magnetic field strength of $B\leq6\times10^{8}$ G at the magnetic poles. Using the same coordinated \nustar{} observation, \citet{2019ApJ...873...99L} estimated an upper limit on the magnetic field strength which is $\le6.7\times10^{8}$ G. Moreover, if we assume $k_{A}=0.5$ \citep{2005ApJ...634.1214L}, then the magnetic field strength at the poles would be $B\leq2\times10^{9}$ G.\\

\section*{Postscript}
After completing this manuscript, it has come to our notice that the same coordinated \nustar{} observation (Obs. ID: $30363001002$) of this source along with some other sources have also been analysed by \citet{2019ApJ...873...99L}. Indeed, submission of the initial version of this manuscript \citep{2019arXiv190202190M} to the arXiv and to this journal postceded the appearance of their paper by only a few hours. Here we carefully analysed their work and compared our result with theirs as suggested by the honorable referee.

\section{Acknowledgements}
We thank the anonymous referee for their critical comments which have improved the content of the paper considerably. This research has made use of data and/or software provided by the High Energy Astrophysics Science Archive Research Centre (HEASARC). This research also has made use of the \nustar{} data analysis software ({\tt NuSTARDAS}) jointly developed by the ASI science center (ASDC, Italy) and the California Institute of Technology (Caltech, USA). ASM would like to thank Inter-University Centre for Astronomy and Astrophysics (IUCAA) for hosting him during subsequent visits. BR also likes to thank IUCAA for their hospitality and facilities extended to him under their Visiting Associate Programme. We thank Dr. M. Pahari, Royal Society-SERB Newton International Fellow, for his useful suggestions in this work.

\def\apj{ApJ}
\def\apjl{ApJl}
\def\pasp{PASP} \def\mnras{MNRAS} \def\aap{A\&A} \def\physerp{PhR} \def\apjs{ApJS} \def\pasa{PASA}
\def\pasj{PASJ} \def\nat{Nature} \def\memsai{MmSAI} \def\araa{ARAA} \def\iaucirc{IAUC} \def\aj{AJ} \def\aaps{A\&AS}
\bibliographystyle{mn2e}
\bibliography{aditya}

\end{document}